\documentclass[iop]{emulateapj}
\usepackage{graphicx}
\usepackage{hyperref}
\usepackage{natbib}
\usepackage{color}

\bibpunct{(}{)}{;}{a}{}{,}

\shorttitle{AMD in Mrk~509}
\shortauthors{Adhikari et al.}
\slugcomment{Accepted for publication in Astrophysical Journal}

\begin{document}


\title{Absorption Measure Distribution in Mrk~509}


\author{T. P. Adhikari\altaffilmark{1}, A. R\'o\.za\'nska\altaffilmark{1}, 
M. Sobolewska\altaffilmark{1,3} and B. Czerny\altaffilmark{2} }
\email{tek@camk.edu.pl}
\altaffiltext{1}{Nicolaus Copernicus Astronomical Center, Polish Academy of Sciences,
    Bartycka 18, 00-716, Warsaw}
\altaffiltext{2}{Center for Theoretical Physics,
    Polish Academy of Sciences, Al. Lotnikow 32/46
02-668, Warsaw}
\altaffiltext{3}{Harvard-Smithsonian Center for Astrophysics, 60 Garden Street, Cambridge, MA 02138, USA}

\begin{abstract}
In this paper we model the observed absorption measure distribution
(AMD) in Mrk 509, which spans three orders of magnitude in ionization 
level with a single-zone absorber in pressure equilibrium.
 AMD is usually constructed from observations of narrow absorption 
lines in radio-quiet active galaxies with warm absorbers. We study the properties of
 the warm absorber in Mrk~509 using recently published broad-band spectral energy 
distribution observed with different instruments. This spectrum is an input in 
radiative transfer computations with full photoionisation treatment using 
{\sc titan} code. We show that the  simplest way to fully reproduce the shape of 
AMD is to assume that the warm absorber is a single zone under constant total pressure. With this assumption we found theoretical AMD which matches the observed 
one determined on the basis of 600 ks RGS XMM-Newton spectrum of Mrk~509. 
The softness of the source spectrum and the important role of the free-free emission breaks
the usual degeneracy in the ionization state calculations, and the explicit dependence of the depths of AMD 
dips on density open a new path to density diagnostic for the warm absorber. 
In Mrk 509 the implied density is of the order of $10^8$ cm$^{-3}$.  
\end{abstract}





\section{Introduction}

There is a general consensus based on high resolution X-ray data that the majority 
of the Seyfert galaxies contain  ionised absorbing gas in their line of sight. 
After the advancements in the observing  X-ray instrumentation as 
{\it Chandra, XMM-Newton, 
Suzaku}, the  extensive studies of the Warm Absorber 
(hereafter WA) in AGN were performed. 
Many narrow absorption lines from highly ionised elements, detected with the use 
of gratings, provided a great opportunity to study this warm material; 
\citep[][and many other papers]{kaspi2001,collinge2001,kaastra2002,behar2003,netzer2003,
krongold2003,yaqoob2003,steen2003,blustin2003,rozanska2004,turner2004,steen2005,
costantini2007,winter2010,winter2012,tombesi2013,laha2014}.

Some physical parameters of WA were estimated, but some are still difficult to constrain.
From energy shifts of line centroids, it was found that the 
absorbing matter is  systematically
outflowing with velocities in the range $10^2-10^3 ~\rm kms^{-1}$ 
\citep{kaspi2001,kaastra2002}.
Directly from observations, ionic column densities of 
particular ions could be found and they are typically of the order of $10^{15-18}$ cm$^{-2}$.
In the next step, by assuming separate ionisation zone for each ion, it is possible to calculate
the equivalent hydrogen column densities, 
knowing ion's ionisation fraction from photoionisation 
codes such as {\sc cloudy, xstar, slab, xabs}. 
Such equivalent hydrogen column densities are in the range of $10^{18-22}$ cm$^{-2}$ 
\citep{steen2005,costantini2007}. Moreover, those densities correspond to the continuous change
of ionization parameter $\xi$ ( see Eq.~\ref{eq:xi} for definition), which spans the range of
 a few decades, log$(\xi) \sim -1$ up to  4.

Since then, several attempts have been made to model continuous ionisation structure of 
the WA in several AGN  \citep[][]{holczer2007,behar2009,detmers2011}.
Furthermore, \citet{holczer2007} proposed to describe ionisation structure of the wind 
by showing absorption measure distribution (AMD) obtained by a derivative formula
(Eq. 5 in this paper). These authors, for the first time, have shown that in the case of 
the source IRAS 13349+2438, AMD obtained  from 
observations has deep minimum in column density that is consistent with the negligible 
absorption of gas with log$(\xi)$ between 0.8 and 1.8. 
Such deep minima are present in AMD of other objects as well 
\citep[see for instance][]{behar2009,detmers2011,stern2014}, and they are interpreted as 
the observational evidence for thermal instability in a given ionisation and  the  
temperature regime. 

To reproduce the observed  AMD theoretically, we should consider continuous ionisation 
structure of WA
in such a way that we can obtain proper normalisation of the AMD function. It has been done
recently by \citet{stern2014} assuming a radiation pressure confinement (hereafter RPC) of the WA
material. The authors considered that the WA consists of the stratified clouds, 
for which $P_{\rm gas}$ increases gradually with decrease of $P_{\rm rad}$. The 
exponential decrease of the radiation pressure is caused via absorption on the 
material when the distance from the illuminating source, i.e. active nucleus, 
gradually increases. 
The authors were successful in  reproducing the normalisation and the slope of AMD for Sy1 
galaxies by showing the comparison of the model with observational points of AMD for the 
six best studied sources. Nevertheless, they were not able to quantitatively 
reproduce the deep minimum  in column density for those 
six objects present in the log$\xi$ between 0.8 and 2.

In this paper, we show how AMD for the  Sy1.5 galaxy Mrk~509 can be successfully 
reproduced by the WA being under  the condition of constant total pressure, 
which is a sum of gas and radiation pressure
\citep{rozanska2006,goncalves2006}. The radiation pressure is computed from the 
radiation field at each point of the computations and it is added to the gas pressure. We neglect 
other pressure components, for instance the magnetic one. 
We used the broad band spectral energy distribution of Mrk~509
 from the paper of  \citet{kaastra2011} (hereafter K11) to represent the radiation 
illuminating the WA cloud. Assuming the plane
parallel geometry, we carried the simulations
using {\sc titan} photoionisation code \citep{dumont2000}.
By assumption of constant total pressure, the matter of the WA is 
self consistently stratified without any additional requirements. The decrease of the 
radiation pressure is self consistently computed within photoionisation code {\sc titan}
and occurs due to absorption and emission i.e. all possible interactions 
of radiation field with matter.  Since exact radiative transfer is computed, radiation 
pressure self consistently influences the matter structure. 
For the differences between escape probability mechanism and exact radiative transfer 
calculations we refer readers to the paper by \citet{dumont2003}.

Since the radiation field influences the matter structure and eventual thermal 
instabilities, S-curve, which is the dependence of the temperature $T$ on the dynamic 
ionisation parameter $\Xi$ (Eq.\ref{eq:bigxi} for definition),
 can be fully constructed with the use of a single cloud and with 
{\sc titan} calculations \citep{rozanska2006}. 
In this paper, we compare results obtained with two codes, {\sc titan} and {\sc cloudy}, 
for exactly the same physical input:  spectral energy distribution (SED)  of Mrk~509
as the incident radiation. 
The two codes differ in the way they treat the radiative transfer, and only with
the use of {\sc titan} we managed to compute thermally unstable regions and fully 
reproduce deep minima in the observed AMD of Mrk~509 \citep{detmers2011}.  

This result is a direct proof that thermal instabilities are responsible for the deep 
minimum in the column densities of AMD. Moreover, it 
provides strong evidence that the WA  in AGN  is compressed by radiation pressure, and 
it should be modelled with proper treatment of this effect.

The overall structure of the paper is as follows: section~\ref{sec:inc}
 contains the information about the incident radiation shape we used in photoionisation 
calculations. The description of the photoionisation codes and results of the 
photoionisation calculations are given in section~\ref{sec:pho}.
In section~\ref{sec:amd}, we present the method of determining AMD from observations
 and calculating AMD from our models. Finally, we compare our 
theoretical AMD with the observational data. The conclusions are presented in section~\ref{sec:dis}.

\section{Mrk~509 and its spectral shape}
\label{sec:inc}

Mrk~509 with redshift of 0.034397 \citep{huchra1993},
is one of the best studied local AGN, with exceptionally high luminosity 
L(1-1000 Ryd)=$3.2 \times 10^{45}$ erg s$^{-1}$   classified as a Sy1 galaxy. 
Usually it is considered to be one of the closest QSO/Sy1 hybrids.   
\citet{peterson2004} have reported that Mrk~509 
harbours a super massive black hole of mass $1.4 \times 10^8$ M$_{\odot}$.

The WA in this source was extensively studied with 600 ks RGS (reflection grating 
spectrometer) on board of {\it XMM-Newton} X-ray telescope 
\citep{detmers2011,kaastra2012}. The few tens of absorption lines from 
higly ionized metals were identified, 
allowing for determination of ionic column densities. Furthermore, equivalent hydrogen 
column densities were calculated for each ion, and absorption measure distribution
 was constructed \citep{detmers2011}. 
Additionally, the upper limits on the location of different ionisation phases 
of the WA were estimated, using the variability method, which 
requires the assumption that any observed changes are caused by changes in the 
ionisation states \citep{kaastra2012}. Those authors reported that observed 
ionisation components span the distance upper limits from 5 to 400 pc. 

\begin{figure}
\includegraphics[width=9.0cm]{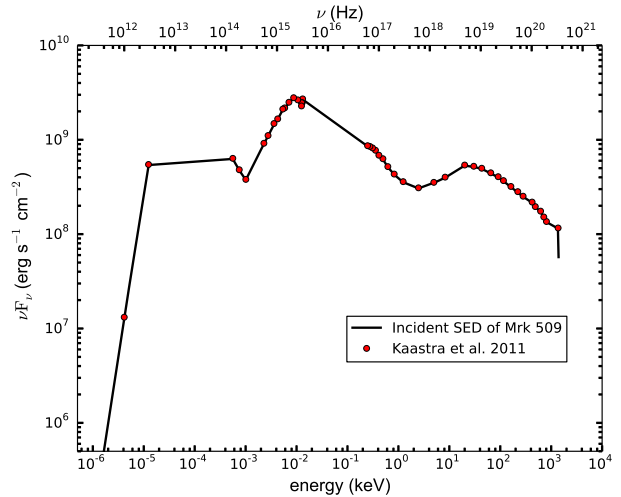}
\caption{\small Spectral energy distribution of Mrk~509. The black line corresponds to the
incident spectrum which is the  input to the photoionisation codes {\sc titan} and {\sc cloudy}
(see text for explanation). The red circles
are the points adapted from K11 and normalised to incident 
flux at the corresponding inner radius of the cloud.} 
\label{fig:incident_sed_std}
\end{figure}

For any consistent  photoionisation calculations of WA, 
the broad-band spectral energy distribution of the source is crucial. 
It was shown by \citet{rozanska2008} that if SED of an AGN is dominated 
by a soft component (i.e. an accretion disk), the stability curve obtained during photoionisation 
calculations differs substantially  from that obtained with the AGN SED dominated by  a hard X-ray 
power-law component. The authors also explained that this is due to the fact that
in the case of soft radiation entering the WA, 
 bremsstrahlung is the dominant cooling in the cloud, while in the case of hard 
X-ray illumination, Compton scattering is responsible for high temperature 
equilibrium in the WA. A systematic study of  how the stability curve depends on the 
shape of the incident SED was done by \citet{cha2009}. 
In particular, the authors discussed the influence of the soft X-ray excess on 
the shape of the stability curve in this paper.
Additionally, chemical abundances affect the ionisation balance of the WA 
\citep{hess1997}, and hence also the shape of the resulting stability curve.
Nevertheless, for this
 paper we assume Solar chemical abundances, and we keep them constant during all
 photoionisation calculations.   
  
Recently, very detailed observed continuum spectral shape of Mrk~509 was published by 
K11. The spectrum was obtained during their multi-wavelength campaign
 of several observations taken in 
various energy bands, with {\it XMM Newton}: RGS, EPIC (pn, MOS), and OM. Simultaneously 
with {\it XMM-Newton} observations, data with {\it INTEGRAL} was also obtained to 
observe the hard X-rays. Additionally, they collected the data of Mrk 509 with 
{\it Chandra} LETGS together with {\it HST} COS UV detection.
Moreover, observation of Mrk 509 was done with {\it Swift}, using both the instruments: 
XRT and UVOT. For the final construction of SED, the authors used IR points from {\it IRAS} 
and {\it Spitzer} data, but it is unclear how much of IR emission is seen by the outflow. 
This is because most probably the wind is located closer to the black hole than the 
dusty torus from which most the IR radiation is originated. 

K11 allowed us to use those points for our project. 
The SED of Mrk~509 covers a wide 
range of wavelength band, which is essential for obtaining the ionisation balance
 needed for photoionisation modelling. All observed spectral points are presented 
in Fig.~\ref{fig:incident_sed_std} as red circles. Mrk~509 has one of the best 
spectra covering broad band among all the well studied AGN.
Black line on the same figure demonstrates the incident SED used by us in photoionisation 
calculations by {\sc titan} and {\sc cloudy} code. Both codes allow for the input of 
incident spectrum by points and the linear interpolation is done in order to fully define 
incident continuum which goes into the simulations. 

\section{Photoionisation Calculations}
\label{sec:pho}
Here, we study the detailed structure of the WA cloud using the phtoionisation codes
 {\sc titan} \citep{dumont2000} and {\sc cloudy} 13.02 \citep{ferland2013} with the 
same initial physical conditions. Although both codes are used to study the clouds 
subjected to the incident radiation emanating from the central source, they do differ in 
their assumptions of radiative transfer treatments and number of spectroscopic lines 
considered. Nevertheless, the general idea is the same. Both codes solve radiation 
transfer through the gas in thermal and ionisation equilibrium with non-LTE 
equation of state. 

The most important difference between codes is that {\sc cloudy} uses escape probability
mechanism to solve radiative transfer, while {\sc titan} uses Accelerated Lambda 
Iteration (ALI) method \citep{collin2004}, which is more accurate in optically 
thick media.   The detailed comparison of both radiative transfer methods in the 
context of X-ray absorbing gas was done in \citet{dumont2003}. Below, we show and compare 
the output of two codes in the case of our particular source Mrk~509 and its warm 
absorber. 

Following assumptions are made for this modelling in both photoionisation codes.
 The geometry of the cloud is set to the plane parallel. 
The expression for the ionisation parameter $\xi$, at the irradiated surface
 of the cloud relating to the distance from the illuminating source $R$ is:
\begin{equation}
 \xi= \frac{L_{\rm ion}}{n R^{2}}
\label{eq:xi}
\end{equation}
where $L_{\rm ion}$ is the luminosity of the ionising source and $n$ is the hydrogen 
number density. 
The luminosity can  be obtained from observations, whereas the number
 density and ionisation parameter are free parameters in our calculations. 

\subsection{{\sc titan} computations}
{\sc titan} was developed by \citet{dumont2000} and was mainly designed to determine 
the structure
 of temperature and ionisation state and the continuum emission spectrum of a thick hot
 photoionised slab of gas. Using this code, the physical state of the gas is computed at
 each depth, assuming the local balance between ionisation and recombination of ions, 
excitation and deexcitation, local energy balance and finally total energy balance. 
{\sc titan} assumes the transfer of both continuum and lines using  
ALI method, which precisely computes line and continuum intensity self consistently. 
There are many options to control the physical properties of the absorbing gas.
Here, to calculate the stability curve
 which is compared to {\sc cloudy} code (Sec.~\ref{sec:stab}), we use the assumption
 of constant density slab, but for the explanation
of AMD in Mrk~509 we reject this assumption and compute more realistic 
constant total pressure cloud  (see results in Sec.~\ref{sec:amd}). This exercise is
made to ensure the reader, that stability curves calculated by both codes agree for 
the same physical conditions of the gas. 

For the purpose of this paper, we make stronger assumption that the total pressure 
(i.e. gas and radiation pressure: $P_{\rm gas} + P_{\rm rad}$) 
is held constant throughout the plane parallel slab of
 the gas, and the volume density and ionisation parameter across the cloud are stratified. 
The assumption of constant 
pressure allows the ionised gas to be naturally stratified due 
to illumination, as discussed in \citep{rozanska2006}. The computations of all the 
structure are done iteratively until the convergence is reached.

In the constant total pressure model, 
plane parallel geometry is considered, and the radiation source 
is assumed to hit the cloud perpendicular to the surface. The parameters of the
 model are the ionisation parameter $\xi_0$, and the hydrogen number density
 $n_{0}$  (both defined at the illuminated cloud surface), 
 the total column density $N_{H}$,  and the incident SED. 
Although the ionisation parameter is well 
constrained in the surface of the illuminated cloud, ionisation properties are continuously
 changing as we go deeper into the cloud. This happens since all thermodynamical parameters 
such as: gas temperature, gas density, and the radiation pressure do change with the 
depth of the cloud.
 To trace this behaviour, the dynamic ionisation
 parameter, $\Xi$ is defined as:
\begin{equation}
 \Xi= \frac{\xi}{4\pi ckT} = \frac{L_{\rm ion}}{4\pi c R^{2}} \frac{1}{nkT} = 
\frac{F_{\rm ion}}{cP_{\rm gas}} = \frac{P_{\rm rad}}{P_{\rm gas}}
\label{eq:bigxi}
\end{equation}
where $F_{\rm ion}$ is a flux affecting the cloud, $c$ is a velocity of light and $T$ is 
the temperature of the gas. 
Here, we do not follow the standard convention that $P_{\rm gas}$ 
accounts only for hydrogen density number $n$. 
For fully ionised gas, with heavy element abundance, the total 
density number equals $\approx 2.3 \, n$, and if the gas pressure is computed 
with total density, the factor 2.3 should be taken into account, as it is in the basic
definition of $ \Xi$,  given by \citet{krolik81}.
Nevertheless, since the observers do not use the value of density while determining AMD from 
observations, we compute the current value of $\xi$ using temperature, radiation pressure and 
the gas pressure, which directly comes out from our code. 

All thermodynamical parameters describing the cloud structure as temperature, 
radiation pressure, hydrogen and total density numbers
are given as functions of the distance from illuminating source. 
Therefore, we can self consistently 
compute both ionisation parameter and its change with the distance from the 
cloud illuminated surface, using the 
value of radiation pressure computed from the second moment of radiation field. 

\subsection{{\sc cloudy} computations}

Cloudy is a well documented public code extensively used in the calculation of the 
photoionisation in many different clouds surrounding planetary nebulas, in the 
interstellar medium and Active Galaxies. It has an extensive explanation of the 
physics of the plasma subjected to different perturbations given in the well organised
 books: (Hazy1 and Hazy2)\footnote{http://nublado.org/}. The same as in {\sc titan} runs 
for {\sc cloudy} calculations, we assumed the 
open geometry with thin plane parallel zones. All other input parameters
as: the incident SED, the ionisation parameter, the inner radius, 
the total column density of the slab and the hydrogen density at the surface of the 
illuminated cloud are the same in both codes. We did this simulation for many grids 
of densities. The default option in the {\sc cloudy} is the constant density through
 the cloud. Nevertheless, there is an option to calculate cloud under constant 
pressure, and we check it and discuss in this paper. 

 \subsection{Stability curves for constant density clouds}
\label{sec:stab} 

\begin{figure}
\includegraphics[width =9cm]{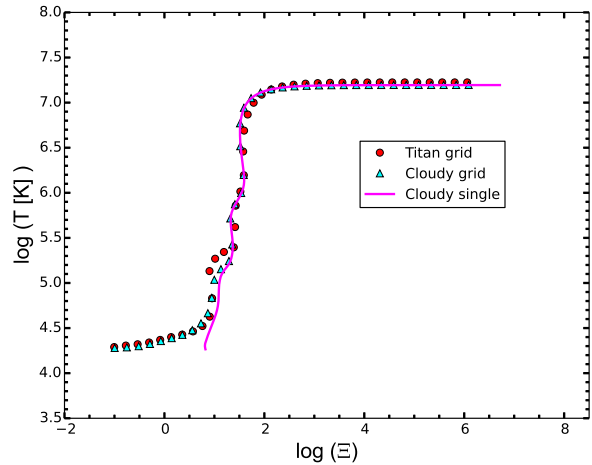}
\caption{\small Equilibrium S-curves for constant density plane-parallel clouds obtained with 
the photoionisation codes {\sc titan} magenta circles, and {\sc cloudy} cyan triangles. 
Additionally, constant density single cloud with $n_H=1$ in spherical geometry computed 
with {\sc cloudy}  is presented by solid magenta line.} 
\label{fig:cd_cooling_curve}
\end{figure}

The general way for the analysis of the stability of the cloud subjected to 
the incident radiation from the central source is to study a stability curve, i.e.
temperature as a function
 of the ratio of the radiation pressure to the gas pressure, which is a dimensionless
 parameter $\Xi$~(Eq.~\ref{eq:bigxi}). Such defined stability curve can be computed in several ways. 
The most natural way is to compute a grid of constant density 
clouds located at the same distance from the illuminated source. By changing the cloud 
density, we change the ionisation parameter and so the gas temperature of the cloud. 
In Fig.~\ref{fig:cd_cooling_curve}, we compare stability curves computed in this way 
by two photoionisation codes {\sc cloudy} (cyan triangles), and {\sc titan} 
(magenta squares).
Both codes use parameter selection of the clouds tabulated in Table ~\ref{tab:param}.
 For each calculation, we considered the source luminosity $3.2\times 10^{45}$ 
ergs s$^{-1}$ and the inner radius of the cloud, i.e., the distance between the source 
and the illuminated face of the cloud is taken as log$(r_0/{\rm cm})=17.25$.

In case of extended envelopes, the radiation pressure changes while we go deeper into the 
cloud due to geometrical dilution and absorption by ionised gas. 
Therefore, in principle, it 
is possible to calculate extended, spherically symmetric cloud at constant density. 
While entering deeper inside the cloud, the luminosity decreases as a distance square, and 
ionisation parameter of the consecutive zone decreases. A full 
stability curve can be reconstructed also in this way, 
as presented in Fig.~\ref{fig:cd_cooling_curve} 
by magenta solid line.  

All curves agree for temperatures down to $3 \times 10^4$ K. Below this temperature, 
atomic data which are different in the two codes may cause some inconsistencies. 
But since we concentrate here on the warm absorbers, we may consider that both 
codes work perfect in the considered temperature regime. 
 
\begin{table}
\caption{\small Constant Density Cloud parameters. When the range of densities is shown 
that means the grid of clouds were calculated. The inner cloud radius in all cases is 
log$(r_0/{\rm cm})=17.25$. }
\label{tab:param}
\begin{tabular}{cccc}
\hline 
Codes Used &$n_0$ (cm$^{-3}$) &$\xi_0$ (ergs cm s$^{-1}$) & Geometry \\
\hline
{\sc titan}  &10$^2$-10$^{12}$&     ~as $L/nr^2$& plane parallel\\
{\sc cloudy} &10$^2$-10$^{12}$&     ~as $L/nr^2$& plane parallel \\
{\sc cloudy}  &1&10$^{11}$& Spherical\\
\hline 
\end{tabular}
\end{table}
\subsection{Stability curves for constant pressure clouds}

\begin{figure}
\includegraphics[width=9cm]{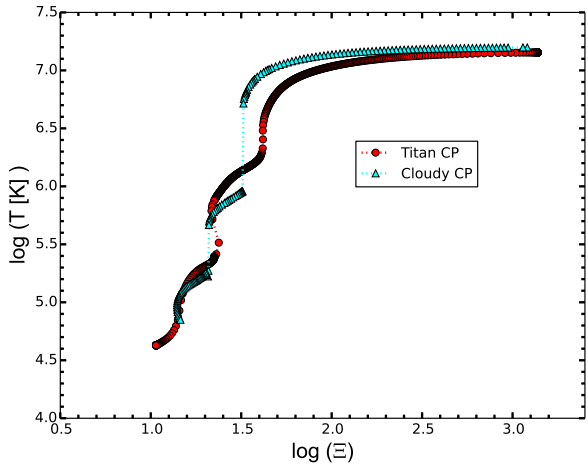}
\caption{\small Equilibrium S-curve obtained from {\sc cloudy} (red) and 
{\sc titan} (cyan).
 Both curves are produced with the number density $n_{0}=10^{6} \rm cm^{-3}$ 
and ionisation parameter $\xi = 10^{6}$  erg cm $\rm s^{-1}$ at the irradiated 
surface of the cloud. The inner radius to the cloud is set to log$(r_o/cm)=16.75$ cm,
to ensure that illuminating luminosity is equal to the observed one.
The total column density ($N_{H}$) of the slab in both calculation is set  to 
$1.5\times10^{23} ~\rm cm^{-2}$. } 
\label{fig:cp_cooling_curve_den1e6}
\end{figure}

As shown in Fig.~\ref{fig:cp_cooling_curve_den1e6}, the constant pressure 
stability curves form 
{\sc titan} and {\sc cloudy} shows the similar trend except for the considerable  
difference in the thermal  instability zone. Thermal instability zones are 
part of the stability curve where the slope is negative. The main difference 
between two codes here is that {\sc cloudy} is not able to compute the structure at 
the zones of instability (there are no points on the unstable branch of stability curve).
The stability curve computed by {\sc titan} shows several points on unstable branch. 
 Also there is a difference in the range of ionisation
 parameter at which the instability occurs. Such difference may be due to the 
different approximations used in the code to solve thermal balance equation and
 the differences in the atomic database used.

\section{Absorption measure distribution}
\label{sec:amd}
The distribution of the absorbing column in the line of sight is 
often described as Absorption Measure Distribution (AMD). Almost always
the large range or ionisation state of gas located on the line of sight towards 
observer is seen in AGN with the warm absorber. A few tens of absorption
lines is detected in high resolution spectra of several AGN. Each line can be treated 
as an constant density absorber and using photoionisation code for the thin slab, 
we can calculate the ionisation parameter and ionic column density $N_{\rm ion}$ of
the slab required to produce such an observed absorbing feature.  
\citet{holczer2007} formulated the expression for 
quantifying the strength of absorption at these different ionisation states as:
\begin{equation}
  {\rm AMD}= \frac{dN_{H}}{d(\log\xi)}
\end{equation}
where, $\xi$ is defined by Eq.~\ref{eq:xi}, and $N_{H}$ is the 
total hydrogen column density along the line of sight.
Furthermore, if solar abundances $A_{Z_\odot}$ are assumed, the equivalent hydrogen 
column density can be calculated separately for each ion using relation:
\begin{equation}
 N_{H} \approx { N_{\rm ion} \over  f_{\rm ion}(\xi_{\rm max})A_{Z_\odot}},  
\label{eq:ion}
\end{equation}
where $f_{\rm ion}(\xi_{\rm max})$ is the fractional ion abundance with respect to the 
total abundance of its element, and $\xi_{\rm max}$ is the maximum value of ionisation parameter 
required to produce the observed feature.
In nature there is a distribution of gas with different $\xi$, which 
may account for absorption in the line, and in reality one must take into account the 
full dependence of $f_{\rm ion}$ on $\xi$.   
This means that when ionic column densities for individual 
warm absorber are measured, the AMD can be derived from the formula:
\begin{equation}
\label{eq:ion2}
 N_{\rm ion} = A_{Z_\odot} \int {\rm AMD} f_{\rm ion}(\log \xi) {d(\log\xi)}.
\end{equation}

It is not possible to get the hydrogen column density 
directly from observation since we do not see hydrogen lines. It is derived using the 
ionic column densities of each ions assuming continuous distribution of the absorbers 
for several objects, including Mrk~509 \citep{detmers2011}. 
It should be noted that the hydrogen column densities 
derived from the observations are nominal and model dependent.
For instance, in \citet{detmers2011} paper, authors always used thin slabs, 
assuming that lines are unsaturated. Even adding the thin slabs together authors make 
a strong assumption that they are on the linear part of the curve of growth, and the 
fitted ionic column densities are always only lower limits.
This kind of analysis is valid if the absorber is optically thin. 
For optically thick lines,
the absorber can posses higher column density with the same strength of saturated lines.
As a result the normalization of AMD fitted by thin slabs can be underestimated 
by one order of magnitude.
Below, we present the 
modelled AMD, computed from the gas with continuous distribution of ionisation 
parameters, as a result of the constant total pressure balance.  

\subsection{Modelled AMD for Mrk 509}
\label{sec:samd}
We used the {\sc titan} photoionisation code to reproduce AMD for Mrk~509 assuming 
that the absorbing cloud is in total pressure equilibrium \citep{rozanska2006} . 
When a cloud is irradiated with the incident SED of the source, 
the total pressure equilibrium requirement self consistently provides the distribution 
of the ionization parameters for which the AMD is calculated.
Note, that when we compute AMD from simulations, we compute $\xi$ at each depth 
by relation:
\begin{equation}
\xi = 4 \, \pi c\, k\, T \, {P_{\rm rad} \over P_{\rm gas(H)}} = 4 \, \pi c\, {P_{\rm rad}
 \over n}.
\label{eq:real}
\end{equation}

 Our model of constant pressure reproduces the AMD derived observationally 
by \citet{detmers2011}. Both discontinuities seen in the distribution of AMD 
in their paper are reproduced using our model. 
We made several exercises how the AMD behaves with different 
parameters and the results are as follows.

\begin{figure}
\includegraphics[width=8.9cm]{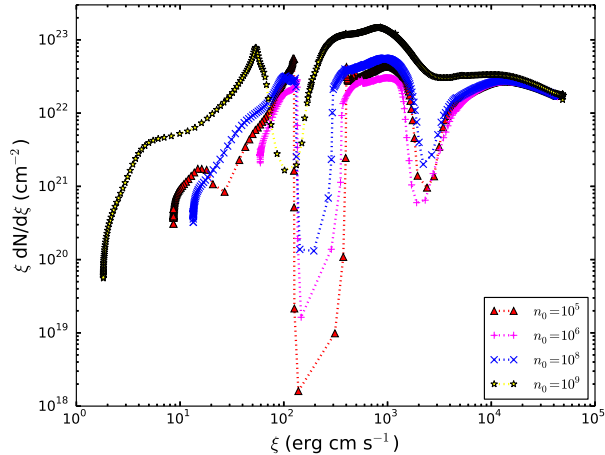}
\caption{\small AMD structure for the same initial ionisation parameter 
$\xi_{0}=\rm 4.3 \times 10^{4}~\rm erg~ cm ~s^{-1}$ and different number densities at illuminated
side of clouds marked in right bottom corner.
}
\label{fig:n_var_xi5}
\end{figure}

\begin{figure}
\includegraphics[width=9cm]{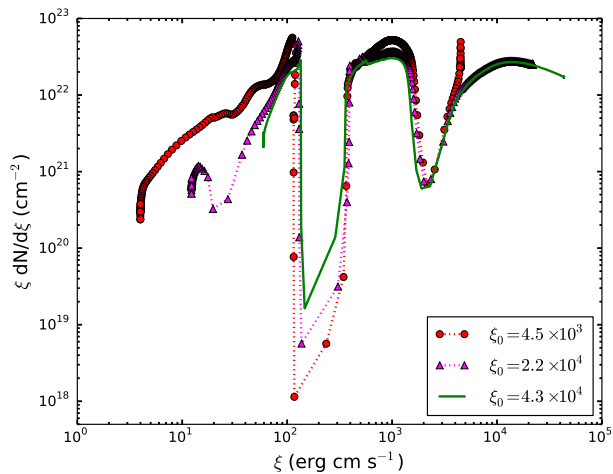}
\caption{\small AMD structure for the same initial number density 
$n_{0}=\rm10^{6}~\rm cm^{-3}$ and different values of ionisation parameters at illuminated side 
of clouds marked in left upper corner.
} 
\label{fig:xi_var_n6}
\end{figure}

For the detail investigation of the effect of the density on AMD structure, 
we made several runs with densities 
$n_{0}= \rm 10^{5},~10^{6},~10^{8} ~and~ 10^{9}~\rm cm^{-3} $ as presented in 
Fig.~\ref{fig:n_var_xi5} (the density  $n_{0}= \rm 10^{7} \rm cm^{-3} $ is 
removed from the figure for clarity). For lower values of the density at the illuminated
face of cloud $n_{0}$ upto $10^{8} \rm~ cm^{-3}$, the position of the dips remains the same . For higher densities 
the position of dips in AMD changes. This is connected with  the fact reported 
by \citet{rozanska2008} that photoionisation models with high density similar to those 
in Broad Line Region (BLR) are not degenerated any more when SED of central source
is dominated by strong optical/UV component. This is a case of Mrk509. The physical reason
for this is that in the case of the hard incident spectrum all radiative processes (Comptonization, line heating/cooling) are linear in density 
and the solution is determined by the ionization parameter, independently from the local density. However, for soft incident spectra the bremsstrahlung plays an important role,
and the quadratic dependence of the bremsstrahlung (free-free) on the density breaks the
degeneracy and introduces a dependence on the 
density \citep{rozanska2008}. 

As can be seen from the Fig.~\ref{fig:n_var_xi5}, when the densities 
are comparable to the Narrow Line Region (NLR) densities, the two dips remain 
prominent almost around the same range of ionisation states. As we move 
towards the regime where the density is comparable to the BLR density, i.e. 
at $ n_{0}=\rm 10^{9}~\rm cm^{-3}$, there is only one 
prominent dip present. Computations for higher densities show only one dip 
in AMD, but the amount of material in the absorber produces much higher normalization of
the modelled AMD than it is observed.
It is difficult to say if the disappearence of this dip is connected with 
a switch in radiative cooling as it is valid in photoionisation models 
of different density. The line cooling may also be important. 
To confirm this result, the different SED should be considered and we plan to 
explore this problem in the future work.  

Overall normalization of the AMD does not change much for densities up to 
$10^8~\rm cm^{-3}$ , but the normalization of dips is different. Dips are deeper for
lower incident densities of the absorber. The depths of the two dips were 
the main criteria used in our comparison of modelled to the observed AMD presented 
below in  Sec.~\ref{sec:holczer}. 

In Fig.~\ref{fig:xi_var_n6}, we show the dependence 
of the AMD structure on the initial ionisation parameter at the irradiated 
side of the absorber, $\xi_0$. It can be seen that the location of 
the instability region remains at the same ionization range inside the absorber.
The difference is only in the normalisation of the dips. 

There is no difference in absorption measure distribution from the 
clouds of the same density affected by various luminosities if we assume the 
same ionization parameter. This is obvious due to the fact that for the same product 
of $\xi n_H$ the same energy flux $L/R^2$ is illuminating the absorber.

\begin{figure}
\includegraphics[width=9cm]{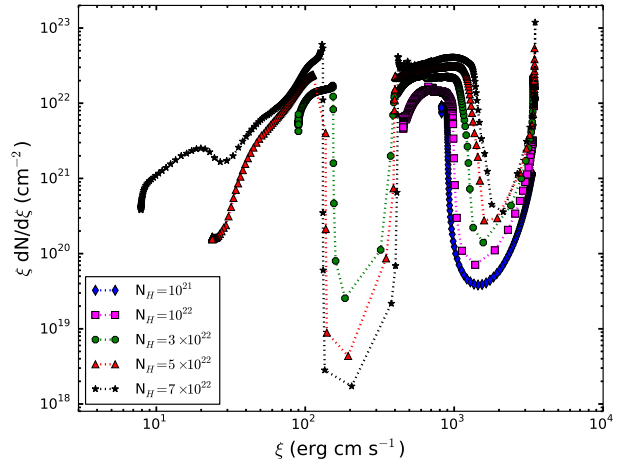}
\caption{\small AMD structure for the same SED and different values of total column density considered.
This is a case with $n_{0}=\rm 10^{8} ~\rm cm^{-3}$ and 
$\xi_{0}=\rm 3.4 \times 10^{3}~\rm erg ~cm~ s^{-1}$. Absorbers with low column densities, less then 
$10^{22}$ cm$^{-2}$, do not have enough matter to pass through two dips in AMD distribution. Those cases
are indicated by blue diamonds and magenta squares.} 
\label{fig:lum_var_n6}
\end{figure}

\begin{figure}
\includegraphics[width=9cm]{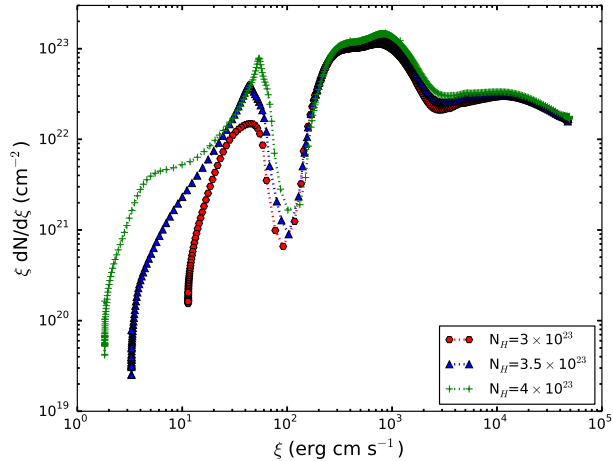}
\caption{\small AMD structure for the same SED and different values of total column denisty
considered. This is a case with $n_{0}=\rm 10^{9} ~\rm cm^{-3}$ and
 $\xi_{0}=\rm 4.3 \times 10^{4}~\rm erg ~cm~ s^{-1}$.} 
\label{fig:lum_var_n9}
\end{figure}

We also investigated the effect of different total column densities of the cloud on AMD. 
It is shown in Fig.~\ref{fig:lum_var_n6} and Fig.~\ref{fig:lum_var_n9} for the number
 densities $n_{0}=\rm 10^8 ~\rm cm^{-3} $ and $n_{0}=\rm 10^9~ \rm cm^{-3} $ 
respectively at the illuminated side the of the absorber.  As we can see 
from the Fig.~\ref{fig:lum_var_n6}, absorbers at lower total column density do not give 
two dips in AMD. Also their overall normalization is lower by
about the half order of magnitude. If the total column density is higher than
 $10^{22}$~cm$^{-2}$, the AMD has always two dips. 
The absorption measure distribution ends at lower ionization parameter for 
higher total column denisty of the constant pressure cloud.
As seen in Fig.~\ref{fig:lum_var_n9}, when the column density is substantial, 
further increase of this parameter does not change the distribution of AMD dips.
This result is consistent with the conclusion made by \citet{stern2014}.

\subsection{Comparison with observations}
\label{sec:holczer}

The direct comparison of an observed AMD with the modelled one is quite hard. 
This is because, for the given source we do observe only from several to several of 
tenth  of absorption lines. Each line puts a point on AMD as described in 
the beginning of Sec.~\ref{sec:amd}. Note, that in case of computed models, 
a few thousands of lines are included, therefore AMD has many more points.
  
To date,  observed AMD are found from individual observations of 
several objects. The general behaviour of AMD in case  of 
five  Sy1 galaxies was given 
by \citet{behar2009}. All of those five galaxies: NGC~3783, NGC~5548, MCG-6-30-15,
NGC~3516, NGC~7469, IRAS~13349+2438, reveal one dip in AMD located more 
or less in the same place between $ \log \xi =0.8-2$. The dip is not explained by 
RPC model presented in this paper. Additionally those objects are rather Sy1 
galaxies with different SED than Mrk~509. The illuminating
SED has the strong influence on AMD. For Sy 1.5 galaxy Mrk~509, AMD reveals two dips 
as shown by \citet{detmers2011}.

\begin{figure}
\includegraphics[width=9cm]{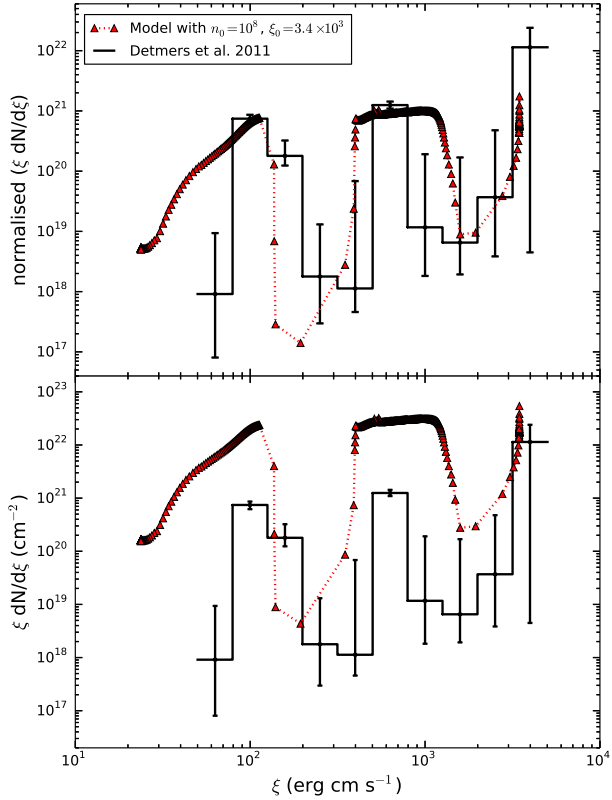}
\caption{\small The comparison of the simulated AMD structure - red triangles, 
 with the observed one - black histogram. Data and error bars are taken from the paper 
by \citet{detmers2011}.
The best model which roughly matches the data is computed for $n_{0}=\rm 10^{8} ~\rm cm^{-3}$,
$\xi_{0}=\rm 3.4 \times 10^{3}~\rm erg~ cm ~s^{-1}$ and $L= 3.2 \times 10^{45}$ erg s$^{-1}$.
Bottom panel presents data with absolute normalization, while in the upper panel 
the modelled AMD is scaled down by a factor $\sim$ 30 to match the observed points on the stable branch. The best fit model represents well the depth of the two dips.} 
\label{fig:c_det-n8-xi3.9}
\end{figure}

The comparison between the observed  AMD for Mrk~509 \citep{detmers2011}  
and our constant pressure model computed with realistic SED input taken from 
current multi-wavelenght studies (K11) is shown in 
Fig.~\ref{fig:c_det-n8-xi3.9}, where observations are given by black histogram,
and red triangles describe our model. Our calculations show the distribution of the 
ionization parameter much more densly. 
Furthermore, our column density presented in the figure by red triangles 
is the realistic column density, with radiation transmitted through all zones in 
the sequence, instead of transmission through a collection of independent zones, 
as in data modelling by \citep{detmers2011}, as given above by Eq.~\ref{eq:ion2}. 
The lower panel of Fig.~\ref{fig:c_det-n8-xi3.9} shows absolute values of AMD 
normalization. It is obvious that modelled normalization is by an order of 
magnitude higher than  the observed one in the stable zones. 
In the upper panel, we re-scale the 
normalization to show that dips are fully reconstructed by our model for the
given incident density and  ionization parameter. We argue here, that the difference in 
the normalization occurs, since observed lines are saturated. Therefore, the derived 
ionic column densities from observations using thin slabs give only lower limits,
 and in reality can be one order of magnitude higher. To check our prediction we have to 
use saturated models to fit the data. They are not very common in the literature. 

There are few values of AMD normalization of several objects reported in the recent papers both from observation and modelling. 
Summarizing, the agreement in the overall shape of the modelled and observed AMD dependence on the ionization parameter is very encouraging but the normalization of the measured and modelled AMD remains as an issue. The observed AMD normalization of Mrk 509 corresponding to the ionization parameter 
$\xi=8\times10^{2}~\rm erg ~cm ~s^{-1}$ is $\sim 10^{21}~\rm cm^{-2}$ \citep{detmers2011}, less by a factor of $\sim$ 30 than the AMD normalization from our best fitted 
constant total pressure model with {\sc titan} ($\sim 3\times 10^{22} ~\rm cm^{-2}$). The AMD modeled by \citet{stern2014} within the frame of their RPC model were shown to be fairly universal, only weakly dependent on the model parameters, and the obtained normalization values were 
$\sim$ $1.1\times10^{22} ~\rm cm^{-2}$. Measurements of AMD in six other objects \citep{behar2009}
implied values of the order of $\sim$ $4\times 10^{21} ~\rm cm^{-2}$. Thus modelled values of AMD normalizations 
are frequently higher than the observed ones, and it is not clear whether the problem lies in the measurement
method or in the modelling.

\begin{figure}
\includegraphics[width=8.5cm]{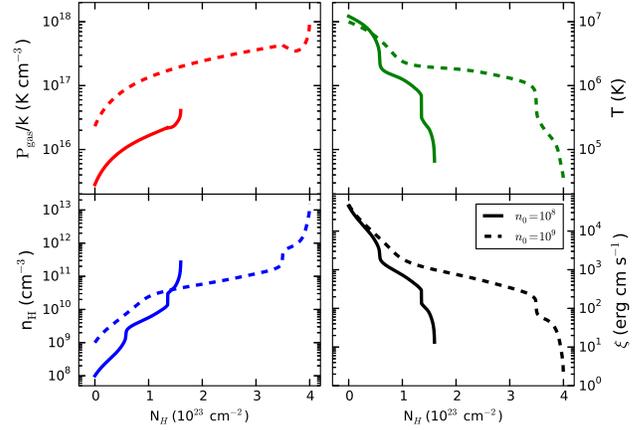}
\caption{\small The structure of pressure (upper left), temperature (upper right), density number 
(bottom left), and 
ionisation parameter (bottom right), versus cloud column density for initial  parameters 
$n_{0}=\rm 10^{8} ~\rm cm^{-3}$, $\xi_{0}=4.3 \times 10^{4}~\rm erg~ cm ~s^{-1}$ (solid lines) 
and $n_{0}=\rm 10^{9} ~\rm cm^{-3}$, $\xi_{0}=4.3 \times 10^{4}~\rm erg~ cm ~s^{-1}$ (dashed lines). In both cases illuminating luminosity is $L= 3.2 \times 10^{45}$ 
erg s$^{-1}$.}
\label{fig:structure}
\end{figure}

Nevertheless, for the first time, the AMD dips can be 
shown in model computations. 
The discontinuities in AMD distribution 
are in the range of the ionisation parameter, $ \log \xi =2-3$ and $\log \xi= 3-4$.
These ranges of $\xi$ corresponds to the ranges of temperatures $T = (4-9) \times 10^5 $ and 
$T=(2-5) \times 10^6$ K, presented in Fig.~\ref{fig:structure} by the green solid line.
This is a proof that the absorbing material should be in pressure equilibrium 
and thin thermally unstable regions can be clearly visible. 
This comparison clearly shows that the warm absorber in Mrk~509 is the continuous
cloud under constant total pressure. 

\begin{figure}
\includegraphics[width=8.5cm]{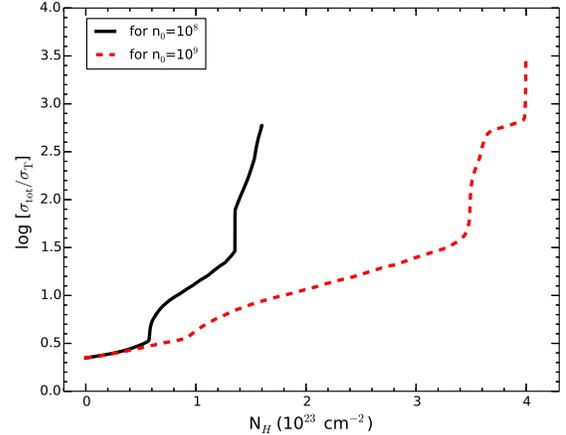}
\caption{\small Total cross section ($\rm \sigma_{tot}$) relative to Thomson cross section
 ($\rm \sigma_{T}$) as a function of cloud column density. The initial cloud parameters are as 
in Fig.~\ref{fig:structure}.}
\label{fig:opacity}
\end{figure}

The two dipps are the result of the thermal instability operating in the irradiated medium. 
To show this effect we present the structure of main cloud parameters as the gas pressure, 
temperature, density and the ionisation parameter $\xi$ as the function of 
column density in Fig.~\ref{fig:structure}. 
We show the plot of total cross section ($\sigma_{tot}$) versus column density in 
Fig. ~\ref{fig:opacity} for the same case of cloud as in Fig.~\ref{fig:structure}. 
Two cases of a sudden rise of the total cross section and a drop in the tempature can be clearly 
noticed, and we argue that they are responsible for the two dips in the observed AMD for Mrk~509. 
The discontinuity is not well resolved in the radiative transfer codes. 
As it was pointed out by \citet{rozanska96} thermally unstable zones are geometrically 
very thin and the correct temperature profile in the transition zone can be recovered 
only after including the
electron conduction. Neither {\sc titan} nor {\sc cloudy} have this option. 
However, the position of the discontinuity is well determined on the basis of 
radiative transfer only, 
as was shown by \citet{czerny2009} and its extension depends on the model parameters. 

Our  result puts strong constrains that 
the warm absorber is build from the material of the density of the order of 10$^8$ cm$^{-3}$.
For SED observed in Mrk~509, the value of density changes the nature of the AMD. 
For densities smaller than $10^9$ cm$^{-3}$, we see two dips of different depth, as observed, and 
the best models compared to the data is for  $10^8$ cm$^{-3}$. For densities 
of the order of  $10^9$ cm$^{-3}$ one dip for higher ionisation parameter is substantially reduced. 
For the given SED we were unable to converge computations for higher densities, but 
we plan to check it for objects from \citet{stern2014} paper, with different shape of SEDs.
Densities much lower than $10^8$ lead to significant overprediction of the depth
of $\xi = 200$ feature, as seen from Fig.~\ref{fig:n_var_xi5}. The discrepancy between the true
modeled AMD normalization and the observed normalization, caused by the line saturation problems
possibly broadens the acceptable density range. If indeed the stable parts of AMD are underestimated 
in the data, but the dips are not, the true depth of the dips should be larger than in the data. 
Assuming the true depth of the low $\xi$ dip equals four decades, then the densities as low as $10^6$ 
(but not lower) provide an acceptable solution for this part of AMD. However,
the high $\xi$ dip is then not so well fitted. 

We note, that RPC model presented by \citet{stern2014} should reproduce the observed AMD dips since 
thermal instability is clearly seen in constant pressure cloud presented in
Fig.~\ref{fig:cp_cooling_curve_den1e6}
by cyan triangles. It was shown that {\sc cloudy} code fully reproduces thermal instabilities under 
different circumstances \citep{rozanska96,cha2012}. In our opinion, the lack of AMD dip in 
the \citet{stern2014}  RPC model is due to the final binning of the AMD used by those authors. 
They have used final bins of 0.25 dex 
in $\xi$ which is too rough to show thin, unstable zones.  Several very small dips are actually 
present in the RPC model by \citet{stern2014} (see their Fig. 2), 
but none of them explains observations.

\section{Conclusions}
\label{sec:dis}

We performed photoionisation simulations of the warm absorber as a cloud  illuminated
by the energy emitted from the AGN centre. Computations were done 
assuming total pressure, $P_{\rm gas}+P_{\rm rad}$ to be constant, and with the use 
of {\sc titan} code by \citet{dumont2000}.
 
We computed absorption measure distribution for the case of Mrk~509, for 
which spectral energy distribution is well known from the multi-wavelength long
term monitoring. 
We found discontinuities in AMD in 
the range of the ionisation parameter,$ \log \xi =2-3$ and $\log \xi= 3-4$, 
similar to that obtained observationally by \citet{detmers2011}. Such
observed discontinuity was often interpreted as the absence of 
of ions in the thermally unstable regions 
\citep{holczer2007}, which can also be seen in the cooling curve in 
Fig.~\ref{fig:cp_cooling_curve_den1e6}. These observed minimum is also 
described as two geometrically distinct regions along the line of sight 
representing high ionisation region and low ionisation region 
\citep{holczer2007}. Contrary to the work by \citet{detmers2011}, instead of two discrete zones, 
we explain AMD as a continuous absorber under constant total pressure.

Since the purpose of this paper is to reproduce the deep minima 
 observed in the continuous distribution of the AMD, the modelled 
AMD is normalised to the observed one for convenience. The model requires 
higher values of column density to produce the structure all the way down 
to the low ionisation parameter.  
As a result of this, the normalisation of the modelled AMD is higher than 
the observed one. Nevertheless, the observed AMD are constructed with some 
approximations discussed in previous sections.  

The model presented in this paper clearly shows that only for the warm absorber with 
substantial density, we can reproduce AMD in Mrk~509.  
We point out here, that \citet{stern2014} did not obtain 
AMD dips from their RPC model since their modeled AMD was computed with too strong binning.
The number of dips observed in AMD may put constraints on the density of the absorbing material, 
which is a crucial parameter needed to indicate the wind location through the relation 
Eq.~\ref{eq:xi}. 
Our work presents new density diagnostic by matching the model with exactly the same 
dips normalization to the observed AMD. The dips strongly reflect geometrically narrow 
thermally unstable regions. For higher densities, one dip disappears for a considered 
SED of Mrk~509. To confirm this result and to search for a parametrization of the
shape of the AMD dips in terms of the incident $n_0$,
it is necessary to consider a sample of sources
and a range of illuminating SEDs. 
We plan to investigate this issue in the future work.

\acknowledgments
{\small We are grateful to anonymous referee for extremelly helpful comments.   
We want to thank J. S. Kaastra for providing us 
the spectra of Mrk~509 and AMD points, we used in our study. We thank A.M. Dumont and M. Mouchet 
for helpful discussion on {\sc titan} calculations and J. Stern for useful discussion on models.
This research was supported by  Polish National Science 
Center grants No. 2011/03/B/ST9/03281, 2013/08/A/ST9/00795, and by 
Ministry of Science and Higher Education grant W30/7.PR/2013. 
It received funding
from the European Union Seventh Framework Program (FP7/2007-2013) under 
grant agreement No.312789.}



\bibliographystyle{apj}
\bibliography{refs}
\end{document}